# An silicon nanocrystal laser

Dong-Chen Wang[#,1], Chi Zhang[#,1], Pan Zeng[2], Wen-Jie Zhou[1], Lei Ma[1], Hao-Tian Wang[1], Zhi-Quan Zhou[1], Fei Hu[1], Shu-Yu Zhang[*,2], Ming Lu[*,1], and Xiang Wu[*,1]

[1] *Department of Optical Science and Engineering, and Shanghai Ultra-Precision Optical Manufacturing Engineering Center, Fudan University, Shanghai 200433, China*

[2] *Department of Light Sources and Illuminating Engineering, and Engineering Research Center of Advanced Lighting Technology (MoE), Fudan University, Shanghai 200433, China*

[#] *equal contributors*

\* corresponding authors: zhangshuyu@fudan.edu.cn, minglu55@fudan.edu.cn, wuxiang@fudan.edu.cn

**Silicon lasers have been the most challenging element for the monolithic integrated Si photonics** [1-6]. **Here we report the first successful all-Si laser at room temperature based on silicon nanocrystals (Si NCs) with high optical gains. The active Si NC layer was made from hydrogen silsesquioxane (HSQ) that had undergone a phase separation annealing, followed by high-pressure hydrogen passivation. Using the nanoimprint technique, a second-order Bragg grating was made on the active layer as the distributed feedback (DFB) laser cavity. This Si NC DFB laser device was optically pumped with femtosecond pulses. Criteria for lasing action,** *i.e.*, **the threshold effect, the polarization dependence and the significant spectral narrowing of stimulated emission were evidently fulfilled, which indicated the realization of an all-Si laser.**

To date, silicon has become the most promising material for the monolithic integrated photonic circuits with a wide range of scientific and industrial applications [3]. Many Si photonic components such as optical waveguides, optical modulators, photodetectors, have been demonstrated successfully [3]. Nonetheless, silicon lasers have not been achieved yet due to the low efficiency of Si light emission [4-6]. To address this issue, tremendous efforts have been paid and a variety of technical schemes have been proposed [7-16]. The first work on optical gains in Si nanocrystals (NCs) was reported in 2000 [7]. Due to the low optical gains and broadband light emission of Si NCs, to realize an all-Si laser has proven difficult since then [14, 17, 18]. Recently, an alternative approach of fabricating matured III-V compound lasers on Si substrates was proposed [19]. However, for better compatibility with the modern Si techniques, all-Si lasers are still in imperious demand for integrated Si photonics.

Efficient approaches to improving optical gains in Si NCs have been developed, such as hydrogen passivation [16, 20] and ion doping [16, 21]. They are either to diminish non-radiative centers or to enhance the population inversion via a kind of sensitizing process. Another key factor for developing Si lasers is to design and fabricate a high-quality optical resonator. Here, we propose the first all-Si laser based on Si NCs with high optical gains. In the scheme, an active Si NC layer with high spatial density and monodisperse Si NCs was developed, which possessed higher optical gains than those made by traditional methods of phase separation [7, 11, 12, 15, 16]. A prolonged and high-pressure hydrogen passivation was then performed, leading to a full saturation of dangling bonds as compared to the normal pressure hydrogen passivation [15, 16, 20]. This full passivation much further increased the gains, and made them comparable to those of usual semiconductor lasing materials such as GaAs and InP [22, 23]. In terms of the physical properties of the high-gain media, a distributed feedback (DFB) laser cavity was designed and fabricated, which finally made the lasing emission available upon optical pumping with femtosecond pulses.

We first prepared the Si NCs-embedded $SiO_2$ active layer, or briefly, Si NC layer, on a fused quartz substrate. Figure 1a shows a cross-sectional TEM image of the Si NC layer. The dark spots indicate Si NCs. The inset illustrates the structure of the Si NC layer. Figure 1b presents a TEM image with larger magnification for a single Si NC as indicated by the white circle. Diffraction fringes are discernible there. Figure 1c displays the absorption spectrum and PL spectrum for Si NC thin film that has been passivated in high-pressure hydrogen for 120 hours. A Tauc's plot of the absorption spectrum is given in figure 1d, from which an optical band gap of 2.75 eV was derived. The difference between the absorption edge and the PL peak position in wavelength, or the Stock's shift, reflects the relaxation of excited electron to a trapped interfacial state before it transits to the ground state of Si NC, as illustrated in figure 1e [24, 25]. This trapped state is at the interface between the Si NC and its surrounding silicon oxide. To further explore the origin of the PL, the decay curve of the PL emission was measured as shown in figure 1f. This behavior has been usually described by equation 1 [16, 26],

$$I(t) = I_1 exp\left[-\left(\frac{t}{\tau_1}\right)^{\beta_1}\right] + I_2 exp[-\left(\frac{t}{\tau_2}\right)^{\beta_2}], \qquad (1)$$

where $I(t)$ is the PL integral intensities as functions of time $t$. $I_1$ and $I_2$ are different intensities of two radioactive recombination at $t = 0$. $\beta_1$ and $\beta_2$ are dispersion factors between 0 and 1. By fitting the decay curve in figure 1f, the two decay times of PL emission, *i.e.*, $\tau_1$ and $\tau_2$, were derived as 7.0 μs and 91.6 μs, respectively. These decay times fall in the range for the transition between the interfacial state and the ground one [7, 24-26], as indicated by figure 1e. *In the following, all the samples refer to those that have been passivated for 120 hours, unless otherwise specified.*

The gain of the Si NC layer was measured by means of variable slit length (VSL) [7, 23], as illustrated in figure 2a. To correct the measured gain, optical loss was acquired by means of shifting excitation spot (SES) [26, 27], as illustrated in figure 2b. Figure 2c shows the integral intensity of PL from the edge emission of the sample against the stripe length according to the VSL approach. The intensity of PL of the sample against the position of the excitation spot is also exhibited according to the SES method. The difference curve of the VSL and SES spectra can be fitted by equation 2 to derive optical gains and losses [15, 16, 26, 27]. In figure 2c, the pump power density was 10.1 mW/cm$^2$.

$$I_{VSL}(l,\lambda) - \int_0^l I_{SES}(x,\lambda)dx = I_0(\lambda) \times \left(\frac{e^{G(\lambda)\cdot l} - 1}{G(\lambda)} - \frac{1 - e^{-\alpha_{Tot}(\lambda)\cdot l}}{\alpha_{Tot}(\lambda)}\right) \quad . \quad (2)$$

On the left hand side of equation 2, the first item is the measured side-emitting PL intensity as a function of stripe length $l$ and wavelength $\lambda$. $\int_0^l I_{SES}(x,\lambda)dx$ stands for the integrated SES intensity as a function of $l$ and $\lambda$, where $x$ is the excitation spot position with respect to $l=0$. On the right hand side of the equation, $I_0(\lambda)$ is the emission intensity, $G(\lambda)$ is the net optical gain, and $\alpha_{Tot}(\lambda)$ is the total optical loss due to absorption and scattering. The real gain $g(\lambda)$ equals $G(\lambda) + \alpha_{Tot}(\lambda)$. In fact, both optical gain and loss depend on the pump power density. In figure 2d, the optical gain as a function of pump power density is given. The threshold power density is 4.9 mW/cm$^2$, *i.e.*, beyond 4.9 mW/cm$^2$, the net gain is positive, but below it, only absorption occurs. A positive net gain is a prerequisite for the stimulated emission. It is seen that when the pump power density is sufficiently high, the net gain can be greater than 450 cm$^{-1}$. This gain is already comparable to those of highly luminescent lasing materials such as InP [22, 23]. Some data of gain and loss at pump power density of 10.1 mW/cm$^2$ are tabulated in Table 1. The samples of 0 h, 24 h, 72 h and 120 h stand for those after 0, 24, 72 and 120 hours of hydrogen passivation, respectively.

Besides optical gain, another basic component for a laser is the optical cavity or resonator. In this work, a DFB resonator was chosen as the laser cavity, the feedback of which was provided by Bragg scattering [28]. In a DFB laser, the resonant wavelength should satisfy the Bragg condition, $m\lambda=2n_{eff}\Lambda$, where $n_{eff}$ is the effective refractive index of the active layer and *Λ* and *m* are the grating period and order, respectively. We fabricated a second-order (*m*=2) Bragg grating by directly patterning the active Si NC layer, as illustrated in figure 3a. Figure 3b gives a photograph of DFB sample on quartz. The 20×20 cm$^2$ quartz was jammed by a vernier caliper, and the grating with size of 10×10 cm$^2$ was visible from diffracted green light. Figures 3c & 3d show a top-view and side-view SEM images of the DFB structure. The average height of the groove is 125.6±0.4 nm and the average distance between two adjacent grooves is 505.0±1.3 nm. The period of 505 nm was chosen to achieve strong resonance by matching the resonant wavelength with the emission peak of the gain medium (λ~ 770 nm). Figures 3e & 3f show the calculated the electric field intensity within the grating and the second-order diffraction peak at the resonant wavelength of 770 nm, respectively.

To achieve lasing emission, the DFB laser was optically pumped by femtosecond pulses. Figure 4a shows the detection scheme for the light emission. In figure 4b, PL spectra for different pumping power densities are plotted. It is seen that for the pump power density sufficiently larger than the threshold (4.9 mW/cm$^2$), the spectral width reduces significantly, which is an indication of stimulated emission. Figure 4c depicts the evolution of the spectral width, or the FWHM (full-width at half-maximum), against the pump power density. The data for samples of 0 h, 24 h and 72 h are also included for comparison. It is seen that when the sample is more passivated, or, the optical gain is larger as shown in Table 1, the spectral narrowing becomes more significant. To visualize this point, figure 4d displays the normalized stimulated PL spectra for differently passivated samples at the pump power density of 10.1mW/cm$^2$. Obviously, the larger the optical gain is, the more (or less) the sharp stimulated (or broad spontaneous) component becomes. The results of figure 4c also suggest that with the increasing optical gain, the threshold power density for lasing is continuously reduced. This trends are more clearly reflected by the so-called threshold effect as shown in figure 4e, where the integral intensity of PL spectrum against the pump power density is plotted. The sub-linear relationships start to change to super-linear ones at 4.9, 5.4, 5.7, 8.1 mW/cm$^2$ for the samples of 0 h, 24 h, 72 h and 120 h, respectively. In figure 4f, for the sample of 120 h, at a power density below the threshold, *i.e.*, 3.4 mW/cm$^2$, the PL spectrum does not change much in either intensity or spectral shape when the polarizer is rotated. However, at a power density far beyond the threshold, *i.e.*, 10.1 mW/cm$^2$, one finds a significant change in PL intensity as shown in figure 4g. By scrutinizing the spectra in figure 4g, a slight shift in the spectral peak position is found when the polarizer is rotated. This indicates the existence of multiple lasing modes in the DFB cavity when lasing occurs [28]. Figure 4h gives the polarization dependence curves before and beyond the threshold pump power density, with all the spectral intensities in figures 4f & 4g normalized to that at 0° in figure 4g. Since spontaneous emission has no polarization dependence in contrast to the stimulated one, the results of figures 4f to 4h offer another evidence for the lasing action of our laser device. The results of figures 4b-4h show that the criteria for lasing action in a DFB cavity, *i.e.*, the threshold effect, the polarization dependence and significant spectral narrowing of the stimulated emission [29] are all satisfied, and lasing actions can happen on the all-Si laser device we developed.

In this Letter, we report to develop an optically pumped all-Si laser device. The active medium was made from HSQ with high density of Si NC, followed by prolonged high-pressure hydrogen passivation. Such a Si NC active layer could show very high optical gains upon optical pumping. The laser resonator was a DFB cavity that could be applicable in integrated Si photonics. Lasing characteristics including the threshold effect, the polarization dependence and the significant spectral narrowing of stimulated emission were fulfilled, suggesting the realization of an optically pumped all-Si laser.

**Acknowledgments**

This work was supported by the National Natural Science Foundation of China (51472051,61275178, 61378080), and Shanghai Sailing Program (16YF1400700). The authors would like to thank Dr. Yue Wang for fabricating the silicon master grating, and Professors Jian Sun, Hai-Bin Zhao, Zong-Zhi Zhang, and Yu-Xiang Zheng for experimental assistance. They also thank Sven Friedl from CIOMP of Chinese Academy of Science for editing the manuscript.

**Contributions**

D.C.W. and C.Z. performed the experiments and data acquisition. P. Z. performed the fabrication of the DFB gratings. W. J. Z., Z. Q. Z and F. H. performed part of data acquisition. L. M. performed the hydrogen passivation. H.T.W. performed part of theoretical analysis. S. Y. Z. performed the design and fabrication of DFB gratings and part of data analysis. M. L. conceived the project, designed the experiments including sample preparation and passivation, and data analysis. X. W. designed the laser device structure, designed the experiments and performed theoretical analysis and data analysis. All the authors contributed to the writing of the manuscript.

**Competing financial interests**

The authors declare no competing financial interests.

**Refrences:**

**Figure & Table Legends**

**Figure 1 | silicon nanocrystals film characteristics**. **a**, Transmission electron microscopic image of silicon nanocrystals film. The inset is a schematic of the Si NC sample **b**, High resolution transmission electron microscopic image of a single Si NC. **c**, The absorption and PL spectra of a Si NC sample. **d**, Tauc's plot of the absorption spectrum. **e**, Schematic energy diagram for the light absorption and emission of Si NC. **f**, Decay curve and its fitting curve for PL of the Si NC sample.

**Figure 2 | Optical gain and loss of silicon nanocrystals film**. **a**, Schematic for the variable slit length (VSL) measurement. **b**, Schematic for the shifting excitation spot (SES) measurement. **c**, The measured PL integral intensities against the slit length (VSL), and the distance of the excitation spot (SES). The pump power density was 10.1 mW/cm$^2$. By fitting the difference curve of the VSL and SES, optical gains and losses were acquired. **d**, Measured net gain against the pump power density. A threshold power density of 4.9 mW/cm$^2$ was observed.

**Figure 3 | Fabrication of second-order Bragg grating. a,** Schematic for the process of grating fabrication using room-temperature nanoimprint. **b,** Photograph of the DFB device (green area, 10×10 mm$^2$). **c,** A top-view of scanning electron microscopic image of the grating. **d,** A side-view of scanning electron microscopic image of the grating. **e,** Simulated near field distribution of the electric filed intensity within the grating. **f,** Calculated resonant wavelength for the TE polarization mode.

**Figure 4 | Device performance and lasing generation. a,** Detection scheme for measuring the lasing characteristics. **b,** PL spectra of the 120 h sample under different

pumping power densities. **c,** FWHM of PL spectrum against the pump power density for the samples of 0 h, 24 h, 72 hand 120 h. **d,** Normalized PL spectra for the samples of 0 h, 24 h, 72 h and 120 h, at pump power density of 10.1 mW/cm$^2$. **e,** Normalized PL intensities against pump power density for the samples of 0 h, 24 h, 72 h and 120 h. Different threshold power densities were found. **f,** PL spectra of 120 h at different polarization angle below the threshold power density. **g,** PL spectra of 120 h at different polarization angle above the threshold power density. **h,** Curves of polarization dependence below and above the threshold power density for the 120 h sample.

**Table 1 | Optical gains and losses of H-passivated samples.**

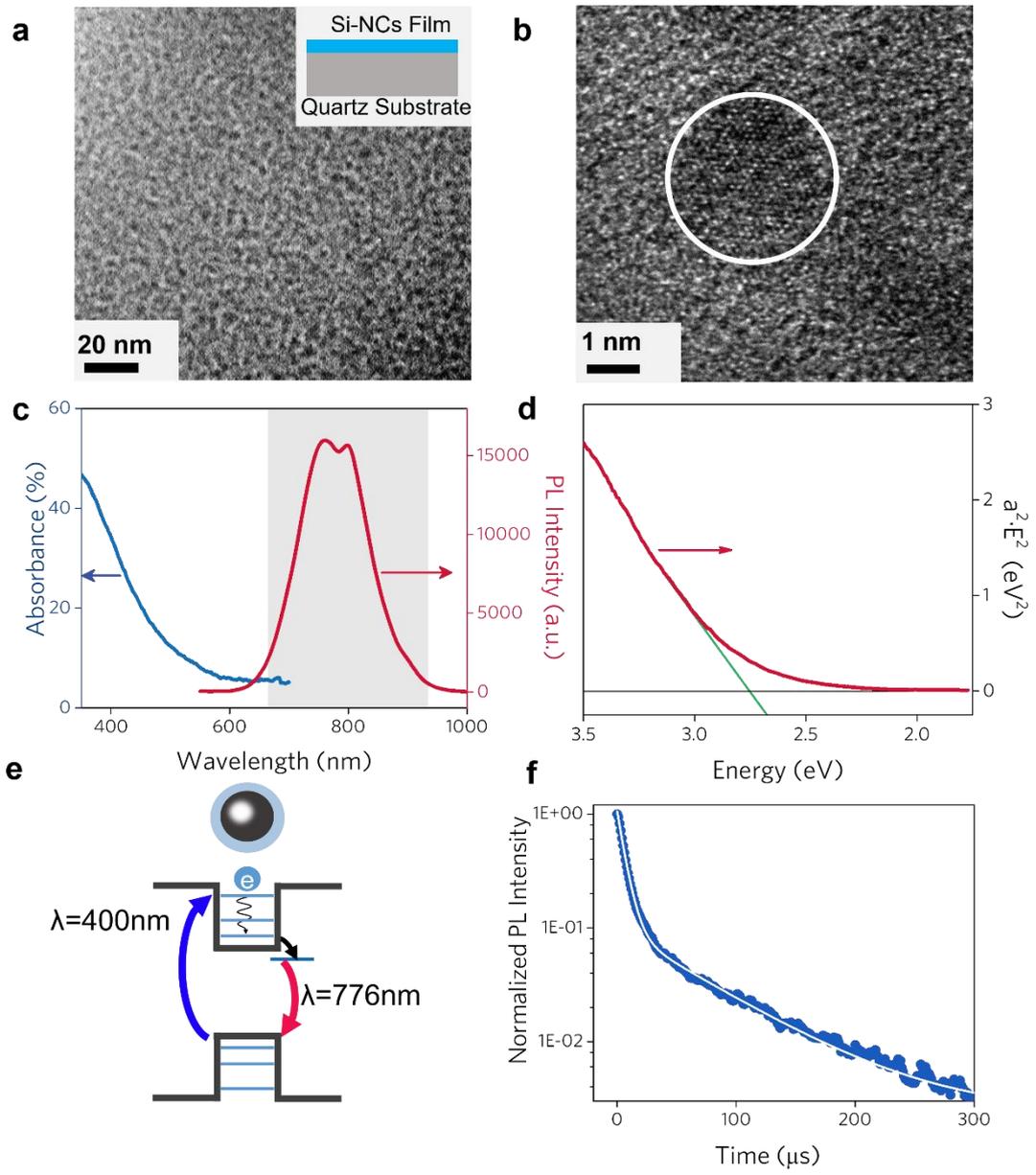

**Figure 1**

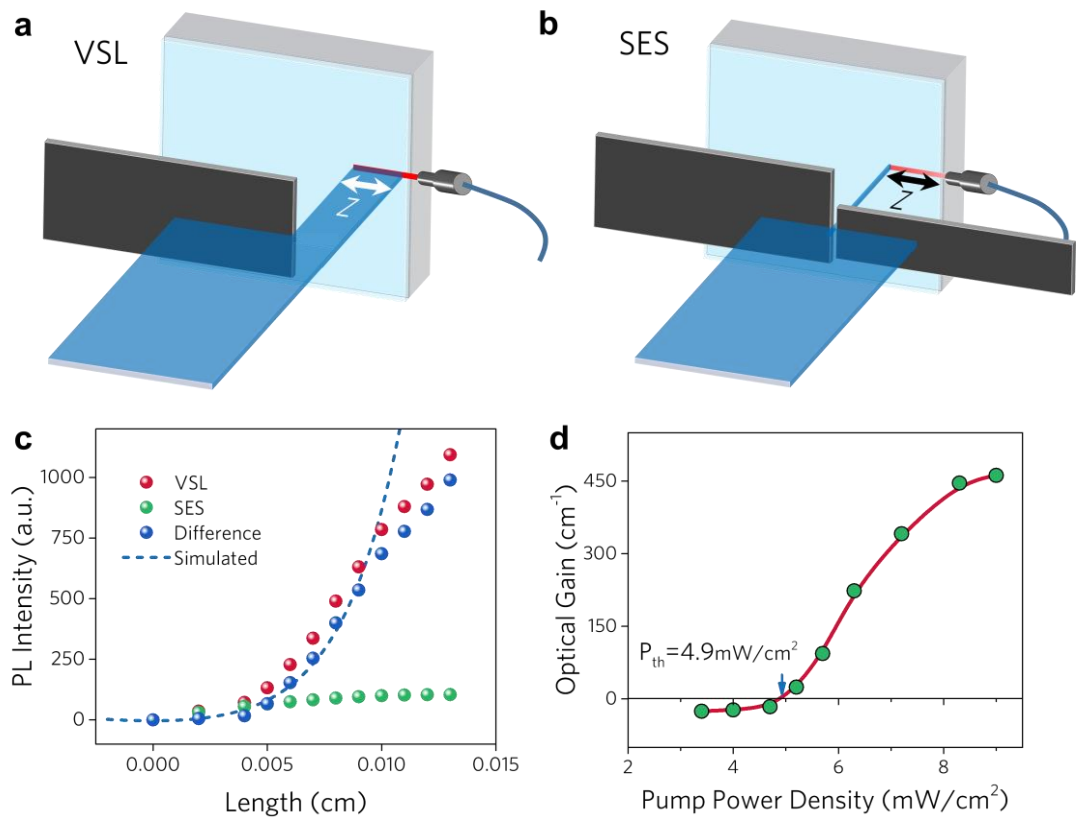

**Figure 2**

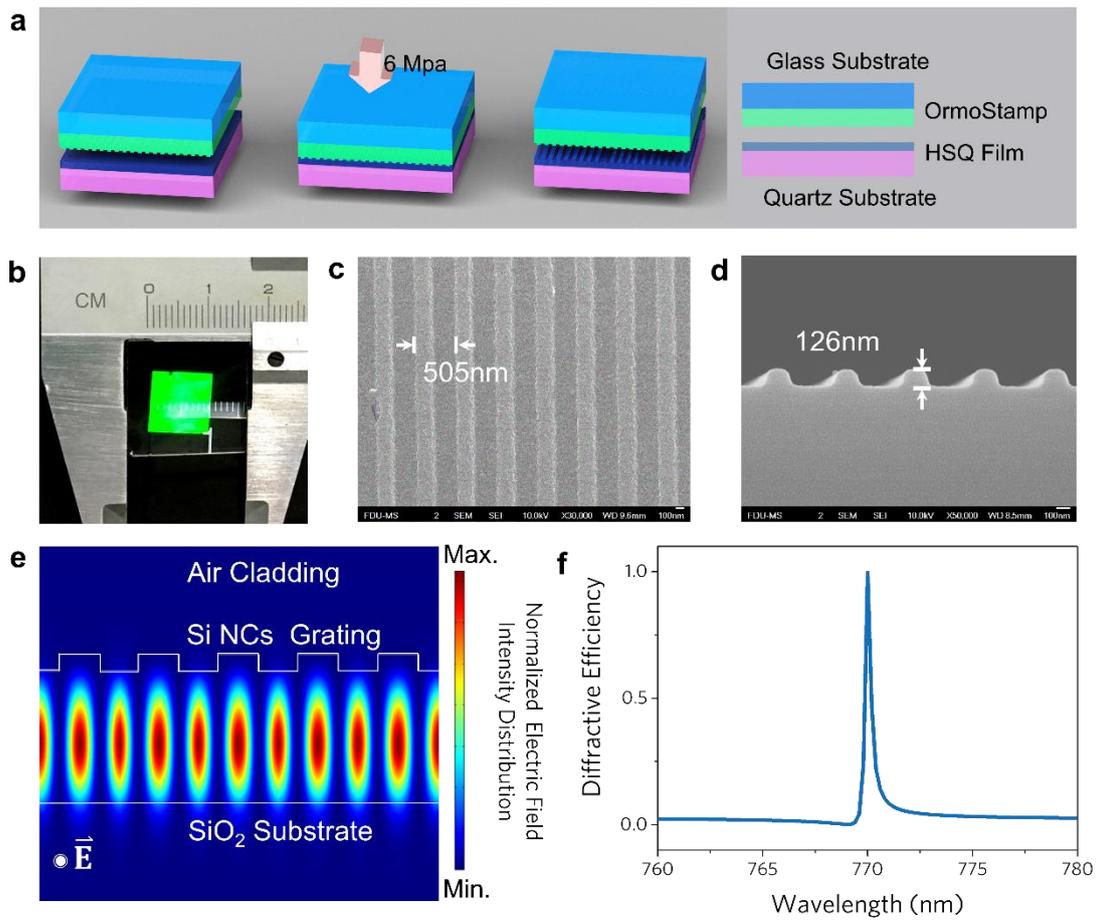

**Figure 3**
16

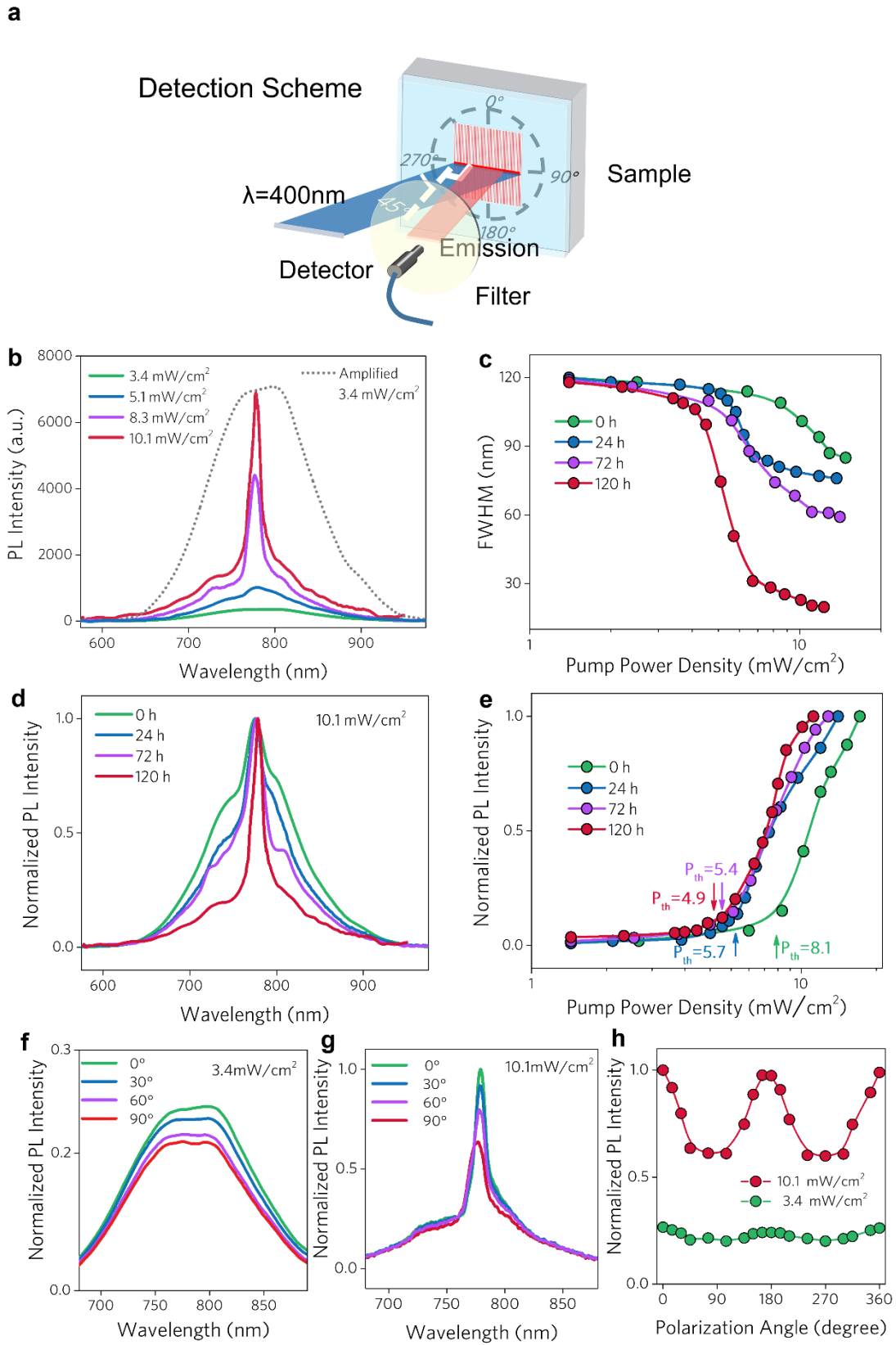

**Figure 4**

### Table 1 | Optical gains and losses of H-passivated samples

| Sample | $G$ (cm$^{-1}$) | $\alpha_{Tot}$ (cm$^{-1}$) | $g$ (cm$^{-1}$) |
|---|---|---|---|
| 0 h | 68.60 | 39.80 | 108.4 |
| 24 h | 260.65 | 36.55 | 297.2 |
| 72 h | 339.41 | 48.45 | 387.86 |
| 120 h | 388.94 | 72.95 | 461.89 |

Table 1